\documentclass[aps,pre,onecolumn,showpacs,groupedaddress]{revtex4}

\usepackage{graphicx,psfrag}

\def\BibTeX{{\rm B\kern-.05em{\sc i\kern-.025em b}\kern-.08em
    T\kern-.1667em\lower.7ex\hbox{E}\kern-.125emX}}

\begin{document}

\title{Determining the minimum embedding dimension for state space reconstruction through recurrence networks}

\author{K. P. HARIKRISHNAN}
\email{kp.hk05@gmail.com}
\affiliation{Department of Physics, The Cochin College, Cochin-682 002, India}
\author{RINKU JACOB}
\email{rinku.jacob.vallanat@gmail.com}
\affiliation{Department of Physics, The Cochin College, Cochin-682 002, India} 
\author{R. MISRA}
\email{rmisra@iucaa.in}
\affiliation{Inter University Centre for Astronomy and Astrophysics, Pune-411 007, India} 
\author{G. AMBIKA}
\email{g.ambika@iiserpune.ac.in}
\affiliation{Indian Institute of Science Education and Research, Pune-411 008, India} 

\begin{abstract}
The analysis of observed time series from nonlinear systems is usually done by making a time-delay 
reconstruction to unfold the dynamics on a multi-dimensional state space. An important aspect of the 
analysis is the choice of the correct embedding dimension. The conventional procedure used for 
this is either the method of false nearest neighbors or the saturation of some invariant 
measure, such as, correlation dimension. Here we examine this issue from a complex network 
perspective and propose a recurrence network based measure to determine the acceptable minimum 
embedding dimension to be used for such analysis. The measure proposed here is based on the well 
known Kullback-Leibler divergence commonly used in information theory. We show that the measure is 
simple and direct to compute and give accurate result for short time series. To show the significance 
of the measure in the analysis of practical data, we present the analysis of two EEG signals as 
examples. 
\end{abstract}


\maketitle

\section{\label{sec:level1}INTRODUCTION}
Nonlinear time series analysis is an important area of applied mathematics with many practical 
applications. The method involves reconstruction of the underlying dynamics \cite {pac} from 
the scalar time series by embedding it in a higher dimension using time delay 
co-ordinates \cite {sau}. Quantifiers from nonlinear dynamics and chaos theory \cite {hil,aba} 
are regularly being employed for the quantitative characterization of the dynamical system 
underlying the time series. For example, the Grassberger-Procaccia (G-P) algorithm \cite {gra} 
is commonly employed for computing one of the most widely used invariant, the correlation 
dimension $D_2$. 

An important aspect in the whole analysis is the identification of the minimum (necessary)  
embedding dimension for reconstructing the dynamics, especially for real world data where 
such information is not known, a priori. If the embedding dimension is less than the actual 
dimension of the system, the computed measures tend to be inaccurate since the dynamics has 
not been completely unfolded. On the other hand, if the embedding dimension used is too large, 
the number of data points in the time series needs to be corespondingly large leading to 
excessive computation. It also enhances the computational error due to the presence of 
additional unwanted dimension where no dynamics is operating.

Two methods are commonly adopted at present to get information regarding the appropriate 
embedding dimension. One method is that of false nearest neighbors (FNN) \cite {ken}. In this 
method, one looks at the behavior of nearest neighbors to a reference point on the attractor 
under changes in the embedding dimension from $M$ $\rightarrow$ $M+1$. The changes in the 
number of nearest neighbors are studied by increasing $M$. When the attractor is unfolded 
completely, the change in the number of nearest neighbors $\rightarrow 0$. 
The second method is to compute some invariant measure, such as, the correlation dimension 
$D_2$ \cite {heg} from time series by increasing $M$. If $D_2$ shows saturation beyond an 
$M$ value, it is chosen as the minimum dimension to compute the nonlinear measures. Both 
methods are efficient and are commonly employed in nonlinear time series analysis, though the 
accuracy of both methods decrease for short time series. 

Over the last one decade or so, a paradigm shift is occuring in the field of nonlinear time 
series analysis, with statistical measures based on complex network theory are increasingly 
being applied for the analysis \cite {mar1,don1}. In this approach, the embedded attractor 
from time series is first transformed into a complex network using a suitable scheme, with each 
point on the attractor identified as a node in the network. If the transformation is done 
properly, one can show that \cite {don2} the structural properties of the embedded attractor 
can be characterized by the statistical measures derived from the complex network. An important 
advantage of this approach is that network measures can be derived accurately from a 
small number of nodes in the network and hence the analysis becomes reasonably accurate even 
for short time series.

The method frequently employed to convert the time series to complex network makes use of the 
property of \emph {recurrence} \cite {eck} of trajectory points in state space. This, again, requires 
an embedding in an appropriate dimension using delay co-ordinates. Nodes corresponding to points on the 
embedded attractor within a recurrence threshold, denoted by $\epsilon$, are considered to be 
connected in the resulting network. The details regarding the construction of the network are 
presented in the next section. We have recently shown \cite {jac1} that the value of $\epsilon$ to be 
used is closely connected to $M$ and hence the knowledge of the correct value of $M$ is important for 
the accurate implementation of the scheme for converting time series to network. So far, a network 
based scheme to estimate the minimum embedding dimension for time series analysis has not been 
proposed in the literature. Our main aim in this work is to propose such a scheme. 

The measure we present is based on the well known Kullback-Leibler divergence \cite {kul1,kul2} used 
to differentiate between two probability distributions, whose details are discussed in the next 
section. The measure is first tested using synthetic time series of known dimension from standard 
chaotic systems and its practical utility is explicitly shown using real world data. Our paper is 
organized as follows: In \S 2, the essential details of the recurrence network construction are 
discussed and the Kullback-Leibler measure is introduced. The measure is tested using time series 
from standard chaotic systems with and without adding noise in \S 3. Practical utility of the 
measure is also presented here using two examples of real world data. Conclusions are 
drawn in \S 4.    

\section{\label{sec:level1}Recurrence networks and the Kullback-Leibler measure}
The first step to compute the required measure is to construct the recurrence network (RN) from the 
time series. RNs are un-weighted and un-directed complex networks and several authors have discussed 
in detail \cite {don1,don3} how to convert a time series to a RN. We have recently proposed a 
scheme \cite {jac1} for this and have used it to study the effect of noise on chaotic attractors 
\cite {jac2}. Here we follow this scheme. Basically, two parameters are involved in the construction, 
a recurrence threshold $\epsilon$ that decides whether two nodes in the network are to be connected 
or not and the embedding dimension $M$. In our scheme, the choice of $\epsilon$ is closely linked to 
the value of $M$. The basic criterion used for the selection of $\epsilon$ is that the resulting 
network just bcomes a single giant component. 

Once the RN is constructed, several statistical measures can be derived from it which are directly 
related to the structure of the embedded attractor. Here we will specifically concentrate on one 
such measure, namely, the clustering coefficient (CC) \cite {bar}. We define the local clustering 
coefficient $C_{\nu}$ of a node $\nu$ as:

\begin{equation}
 C_{\nu} = {{2 f_{\nu}} \over {k_{\nu} (k_{\nu} - 1)}}
 \label{eq:1}
\end{equation}
where $k_{\nu}$ is the degree of the node and $f_{\nu}$ are the number of triangles attached to the 
node \cite {onn}. A triangle is the simplest motif present in a complex network. The value of 
$C_{\nu}$ measures how many of the nodes connected to the node $\nu$ are also mutually 
inter-connected and its value is normalized between 0 and 1. By averaging $C_{\nu}$ for all the 
nodes over the entire network, we get the global CC of the network.

In this work, our main focus is on $C_{\nu}$. We compute the probability distribution 
$P(C_{\nu})$ of the local clustering coefficient of nodes over the entire network. 
The procedure is repeated by increasing 
the embedding dimension $M$ from 2 to 6. We do not go beyond $M = 6$ since we use time series of 
limited length ($N_T < 5000$). We compare the probability distributions for two successive 
$M$ values and check the convergence of the distributions as $M$ increases. In order to  
quantify the difference between two probability distributions, we make use of a  measure 
(denoted as KLM) based on the well known Kullback-Leibler (K-L) divergence \cite {kul1}, 
widely used in information theory \cite {kul2} to differentiate between two probability 
distributions. We compute KLM as a function of $M$ and check its convergence with respect to 
$M$. If the measure shows convergence for $P(C_{\nu})$ beyond $M$, then $M$ is taken as the 
required embedding dimension for the system.

Note that the basic idea behind this procedure is very much analogous to finding the false 
nearest neighbors. When the attractor is embedded in a lower dimension than the actual 
dimension of the system, there will be false neighbors within the recurrence threshold so 
that a reference node gets connected to many nodes which are not real neighbors. Consequently, 
its clustering coefficient $C_{\nu}$ is affected. This is true for all nodes. Beyond the actual 
dimension, the value of $C_{\nu}$ remains genuine, making the probability distribution to 
converge approximately. One advantage of the measure is that it is simple and straightforward 
to compute and can be derived from relatively small number of nodes in the network.

The K-L divergence is usually applied in information theory to differentiate between two 
probability distributions, say $P$ and $Q$. Specifically, the K-L divergence from $Q$ to 
$P$, denoted by $D_{KL}(P|Q)$ is a measure of the amount of information lost when $Q$ is used 
to approximate $P$. For discrete probability distributions, the K-L divergence from $Q$ to 
$P$ is defined as:
\begin{equation}
 D_{KL}(P|Q) = \sum_{i} P(i) \log {{P(i)\over Q(i)}}
 \label{eq:2}
\end{equation}  
For continuous distribution, the summation is replaced by integration. 

Here we compare the probability distributions of the local clustering coefficients $P(C_{\nu})$ for two 
successive embedding dimensions $M$ and $(M+1)$. We then apply the above measure taking $P$ as 
$P(C_{\nu})$ for $M$ and $Q$ as $P(C_{\nu})$ for $(M+1)$ and denote it as KLM:
\begin{equation}
 KLM = |\mu_M - \mu_{M+1}| + \sum_{C_{\nu}} P_M (C_{\nu}) \log {{P_M (C_{\nu})} \over {P_{M+1} (C_{\nu})}}
 \label{eq:3}
\end{equation}
where $\mu_{M}$ is the average value of $C_{\nu}$ for dimension $M$ and $\mu_{M+1}$ is that for dimension 
$(M+1)$. These values are added to capture the difference due to the displacement of one profile from 
the other. The calculation is repeated by taking the distributions for two successive $M$ values at a 
time and plotted as a function of $M$. We find that the measure saturates beyond the embedding dimension 
at which the structure of the attractor unfolds completely. This is taken as the minimum dimension for 
embedding. 

Note that by definition, the measure is always $> 0$, but is not symmetric:
\begin{equation}
 D_{KL}(P|Q) \neq D_{KL}(Q|P)
 \label{eq:4}
\end{equation}
In other words, if one computes the measure from $(M+1)$ to $M$ instead of $M$ to $(M+1)$, then the 
actual value of the measure will be different. However, the final result still remains the same as 
the difference in the measure for successive $M$ values now starts diverging below the actual 
dimension of the system. We have checked and confirmed this. 
We first test the measure using synthetic data from the standard Lorenz attractor 
time series and also data added with different percentage of noise before applying it to real world data. 

\section{\label{sec:level1}Application of Kullback-Leibler measure}
Time series from standard Lorenz attractor is generated with a time step of $0.05$ and data length 
$N_T = 4000$. All analysis in this paper are done with time series of above length to show the potential 
of the proposed measure in the analysis of short time series. The time series is embedded with dimension 
$M$ varying from 2 to 6 with a time delay equal to the first minimum of the autocorrelation. RNs are 
constructed for each embedded attractors using the scheme presented in \cite {jac1}. For each case, the 
probability distribution $P(C_{\nu})$ of $C_{\nu}$ is also computed.

\begin{figure}
\includegraphics[width=0.90\columnwidth]{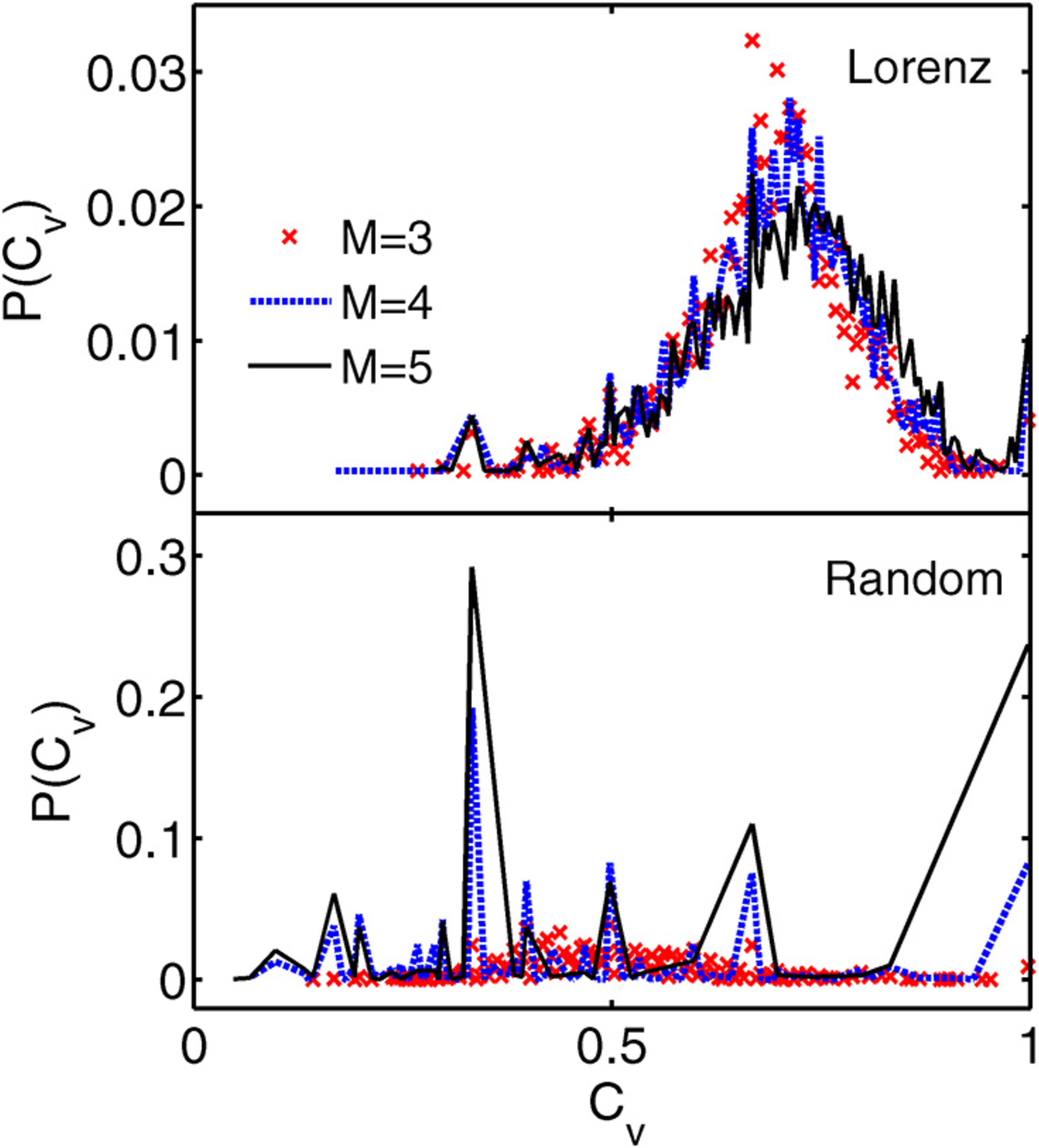}%
\caption{Probability distributions of the local clustering coefficients of nodes of the RNs 
constructed from the time series of the Lorenz attractor (top panel) and that from random time 
series (bottom panel). The results from three different embedding dimensions $(M = 3,4,5)$ are 
shown in both cases.} 
\label{f.1}
\end{figure}
 
The results for $M$ values 3, 4 and 5 are shown in Fig.~\ref{f.1} (top panel). As expected, the three 
probability distributions converge approximately since the dimension of the attractor is $< 3$. On the 
other hand, when the same computation is repeated for random time series, we find that $P(C_{\nu})$ for 
each $M$ vary randomly in the whole interval $[0,1]$ without showing any convergence. The results for 
tha same 3 $M$ values for random are also shown in Fig.~\ref{f.1} (bottom panel).  
In the latter case, the attractor tends to fill the phase space for every $M$ changing the connections 
of each node continuously and hence varying $C_{\nu}$. For a chaotic attractor, on the other hand, 
$C_{\nu}$ and the global CC tend to saturate beyond a certain $M$. 

Another important aspect of the distribution as seen from Fig.~\ref{f.1} is the way in which $C_{\nu}$ 
varies for the two cases. For a chaotic attractor, because of the inherent geometric structure, the 
clustering is generally high and nodes with very low degree or isolated nodes are rare. Consequently, 
nodes with $C_{\nu} \rightarrow 0$ are negligible.  Moreover, the average number of connections for a 
node is also high. For $C_{\nu} \rightarrow 1$, all the nodes connected to node $\nu$ are also to be 
inter-connected, the probability of which is very low. This, in turn, makes the nodes with 
$C_{\nu} \rightarrow 1$ also negligible. On the contrary, for the random RN, the probability for both 
these extremes are comparatively high. Hence, while $C_{\nu}$ for the RN from a chaotic attractor 
generally varies over a small range with the probability distribution having a well defined profile, 
that for random is distributed over the whole interval.  

\begin{figure}
\includegraphics[width=0.90\columnwidth]{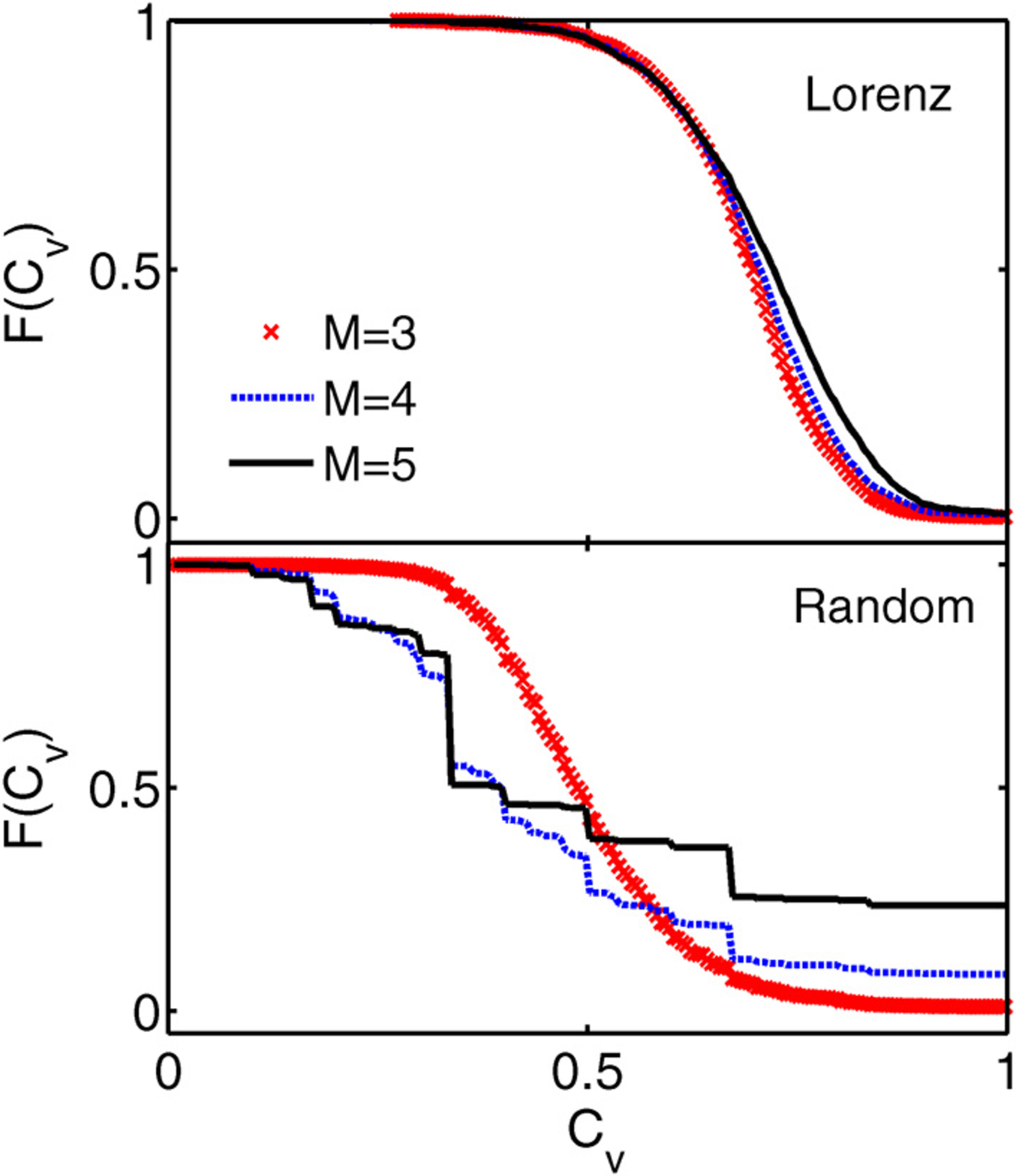}%
\caption{The cumulative distributions (see text) of the probability distributions of the local clustering 
coefficients shown in the previous figure.} 
\label{f.2}
\end{figure}

In order to show the variation of $C_{\nu}$ in a better manner, we compute the cumulative distribution of 
$P(C_{\nu})$ which is generally used to reveal the trend in a probability distribution. It is given by:
\begin{equation}
F(C_{\nu}) = \sum_{C_{\nu}^{'}=C_{\nu}}^{C_{\nu}^{max}} P(C_{\nu}^{'})
  \label{eq:5}
\end{equation}
For the distributions given in Fig.~\ref{f.1}, the cumulative distributions are shown in Fig.~\ref{f.2}. 
Note that the distributions for the 3 values of $M$ for random can be better differentiated in the latter 
figure. We have checked the distributions $P(C_{\nu})$ for RNs from several standard low dimensional 
chaotic attractors and have found that they converge for $M \geq 3$. 

\begin{figure}
\includegraphics[width=0.90\columnwidth]{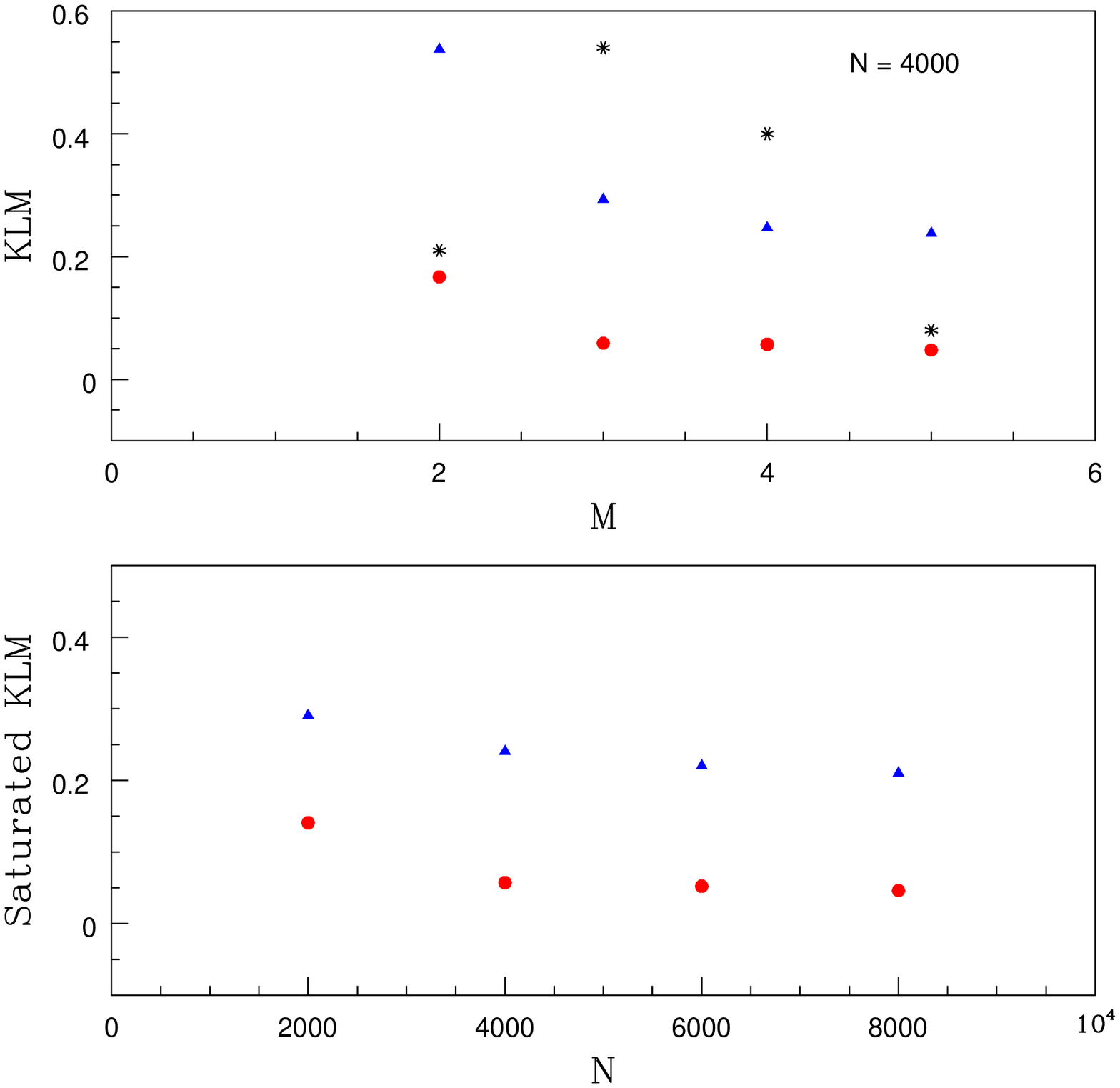}%
\caption{Top panel shows the variation in the Kullback-Leibler measure (KLM) computed with two 
successive embedding dimensions $M$ (randing from 2 to 6) for RNs constructed from the time series 
of Lorenz attractor (solid circle) and the Ueda attractor (solid triangles). The asterisks are the 
corresponding variations of the measure for RNs from random time series. In all cases, the number 
of nodes used $N = 4000$, as indicated. The values of KLM for the RNs from the two chaotic 
attractors saturated upto $M5|M6$ are shown as a function of $N$ in the bottom panel.} 
\label{f.3}
\end{figure}

To quantify this convergence, we now compute the measure KLM taking two distributions with successive 
$M$ values at a time, in all cases. The results for two standard chaotic attractors and random are shown 
in Fig.~\ref{f.3} (top panel). It is evident that the measure correctly identifies the dimension of 
the attractor. We have checked that the saturated values of KLM remains constant with increase in the 
number of nodes $N$ in the RN, which is also shown in Fig.~\ref{f.3} (bottom panel). 

\begin{figure}
\includegraphics[width=0.90\columnwidth]{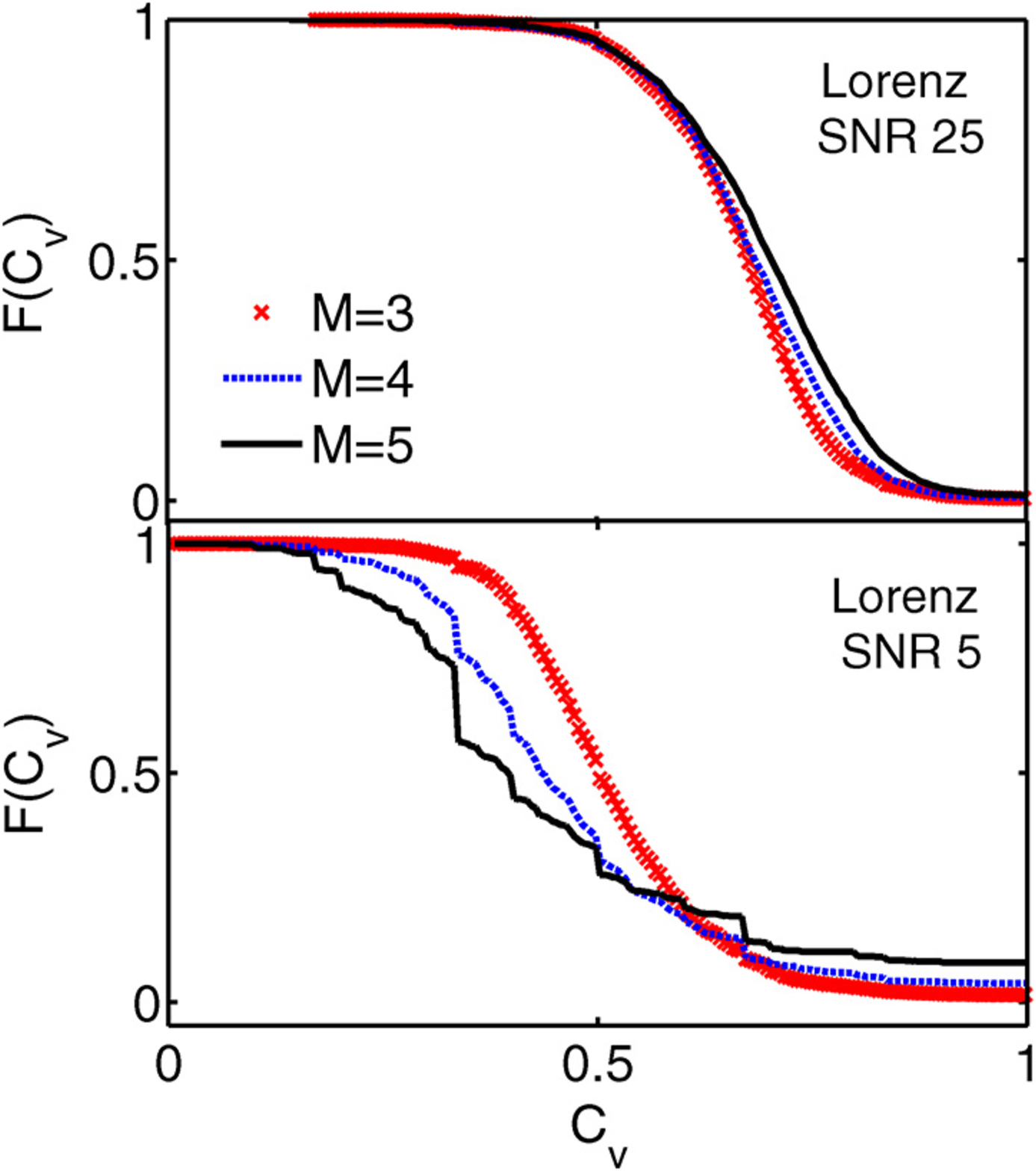}%
\caption{Figure shows how the cumulative distributions of the local CCs from RNs is affected by 
adding noise to a chaotic time series. The top panel is obtained by adding $4\%$ noise while the 
bottom panel is for $20\%$ of added noise to time series from Lorenz attractor. Variation for 
3 embedding dimensions are shown with $N = 4000$.} 
\label{f.4}
\end{figure}

\begin{figure}
\includegraphics[width=0.90\columnwidth]{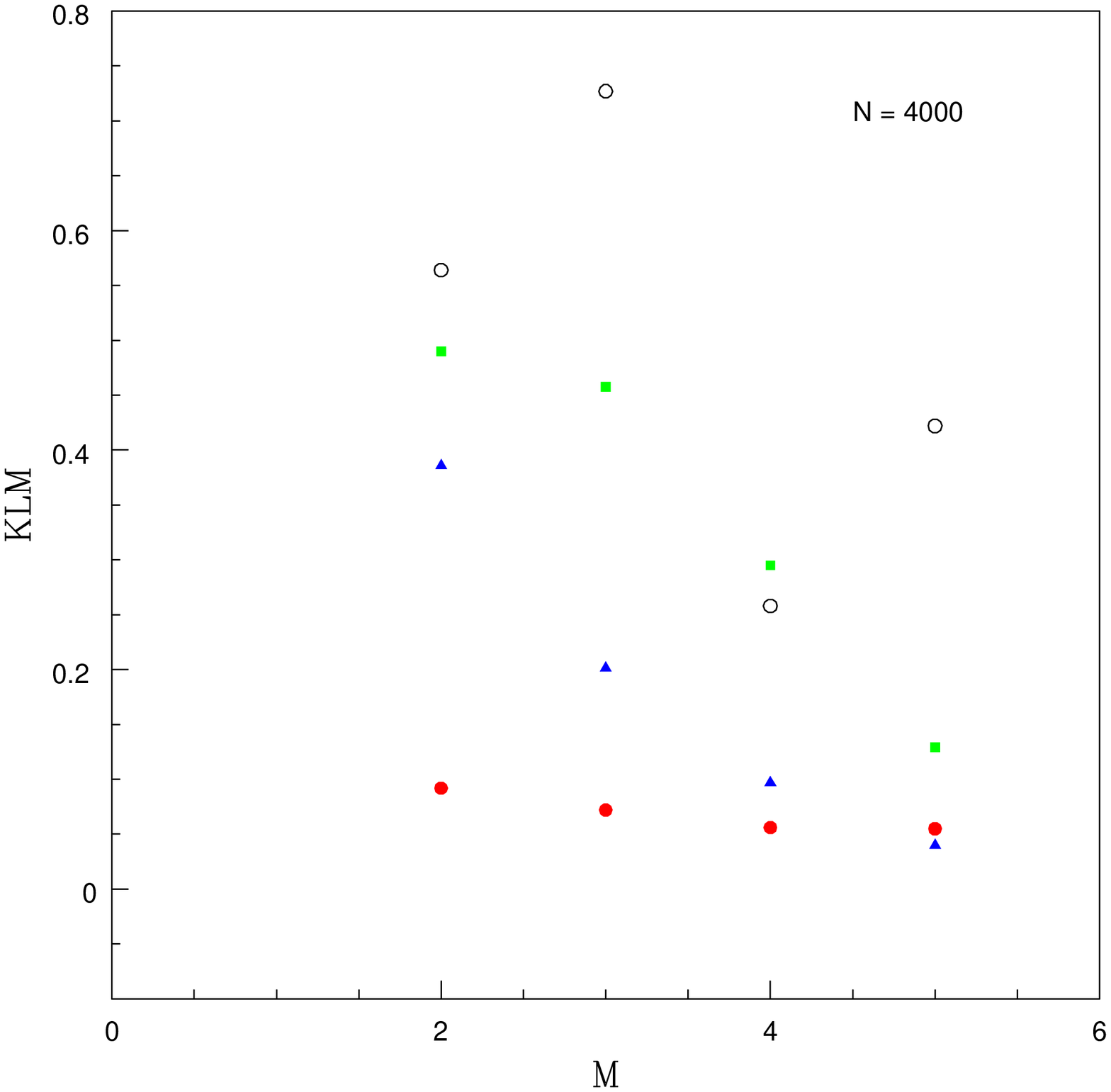}%
\caption{Varition of the measure KLM with $M$ for RNs constructed from Lorenz attractor time 
series with different percentages of noise, namely, $4 \%$ (solid circle), $10\%$ (solid triangle), 
$20\%$ (solid square) and $50\%$ (open circle). In all cases, $N = 4000$. } 
\label{f.5}
\end{figure}

An important issue in the analysis of observed time series is contamination by noise. It is important to 
see how any nonlinear measure is affected by noise added to data, before the measure can be applied to real 
world data. To study the performance of the proposed measure under noisy conditions, we now apply it to 
chaotic data added with different $\%$ of noise. We use the standard Lorenz attractor time series for this 
and generates data by adding $4\%$, $10\%$, $20\%$ and $50\%$ of white noise to it. We undertake the above 
analysis on all these time series by constructing RN and computing KLM. In Fig.~\ref{f.4}, we show the 
cumulative distributions for 3 values of $M$ for time series added with $4\%$ of noise (top panel) 
and $20\%$ of noise (bottom panel). Note that, as the noise level increases, convergence of the distributions 
get shifted to higher $M$ value. To get a better idea of convergence, we compute the KLM as a function of 
$M$ for all the four percentages of noise and the results are shown in Fig.~\ref{f.5}. We find that both the 
actual converged value of KLM and the $M$ value at which convergence occurs increase with $\%$ of noise. 
Moreover, beyond a noise level of $20\%$, the convergence seems to disappear, at least upto the maximum 
$M$ value we have used. In other words, the measure is unable to give a proper embedding dimension if the 
data is contaminated by moderate or high amount $(> 20\%)$ of noise. However, this is true with most 
other measures as well. For example, in the case of FNN, the precise value of $M$ is known to be 
clouded by noise \cite {ken}.     

\begin{figure}
\includegraphics[width=0.90\columnwidth]{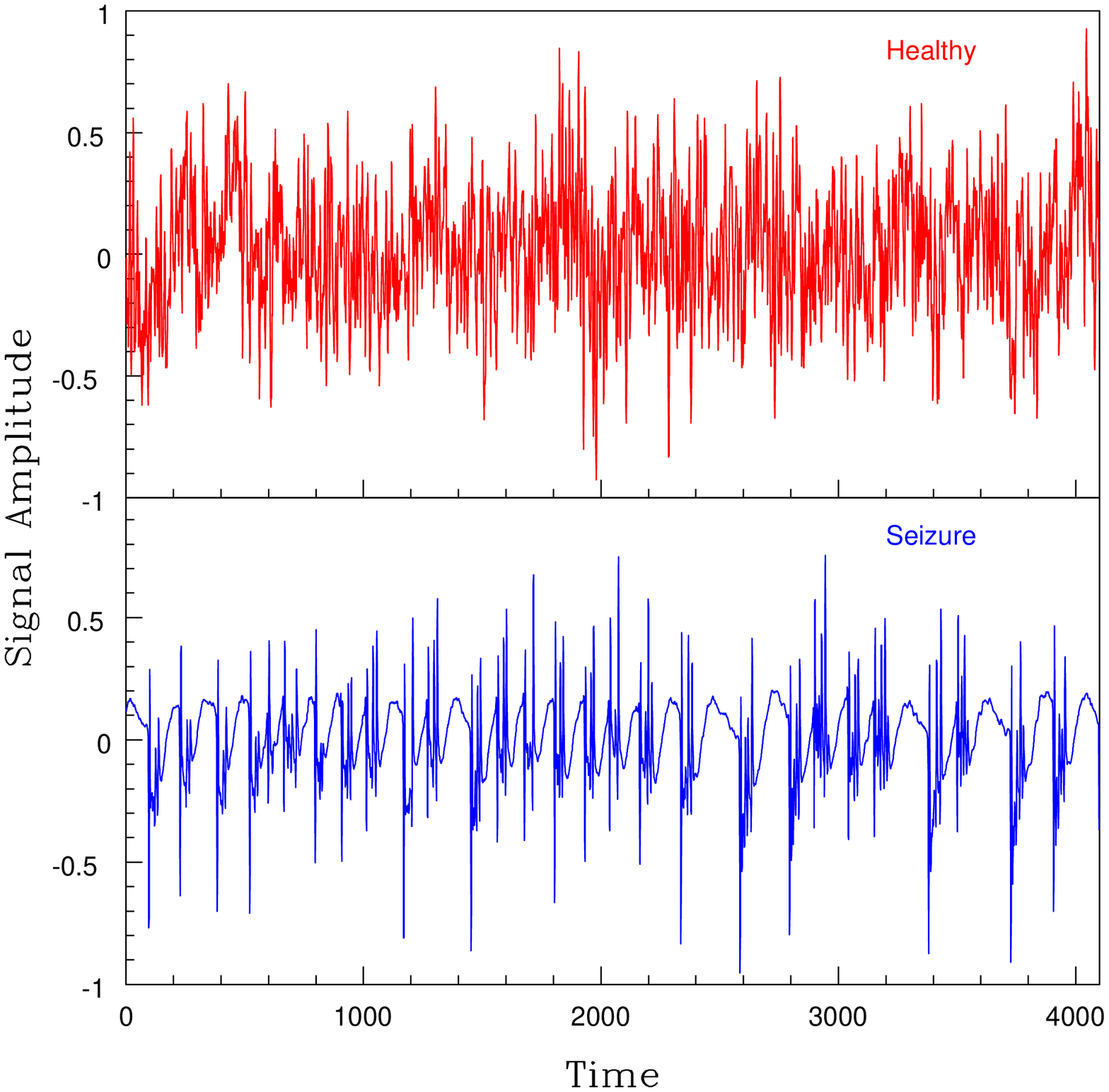}%
\caption{Typical EEG signals (normalized) from a healthy person (top panel) and during 
epileptic seizure (bottom panel).} 
\label{f.6}
\end{figure}

\begin{figure}
\includegraphics[width=0.90\columnwidth]{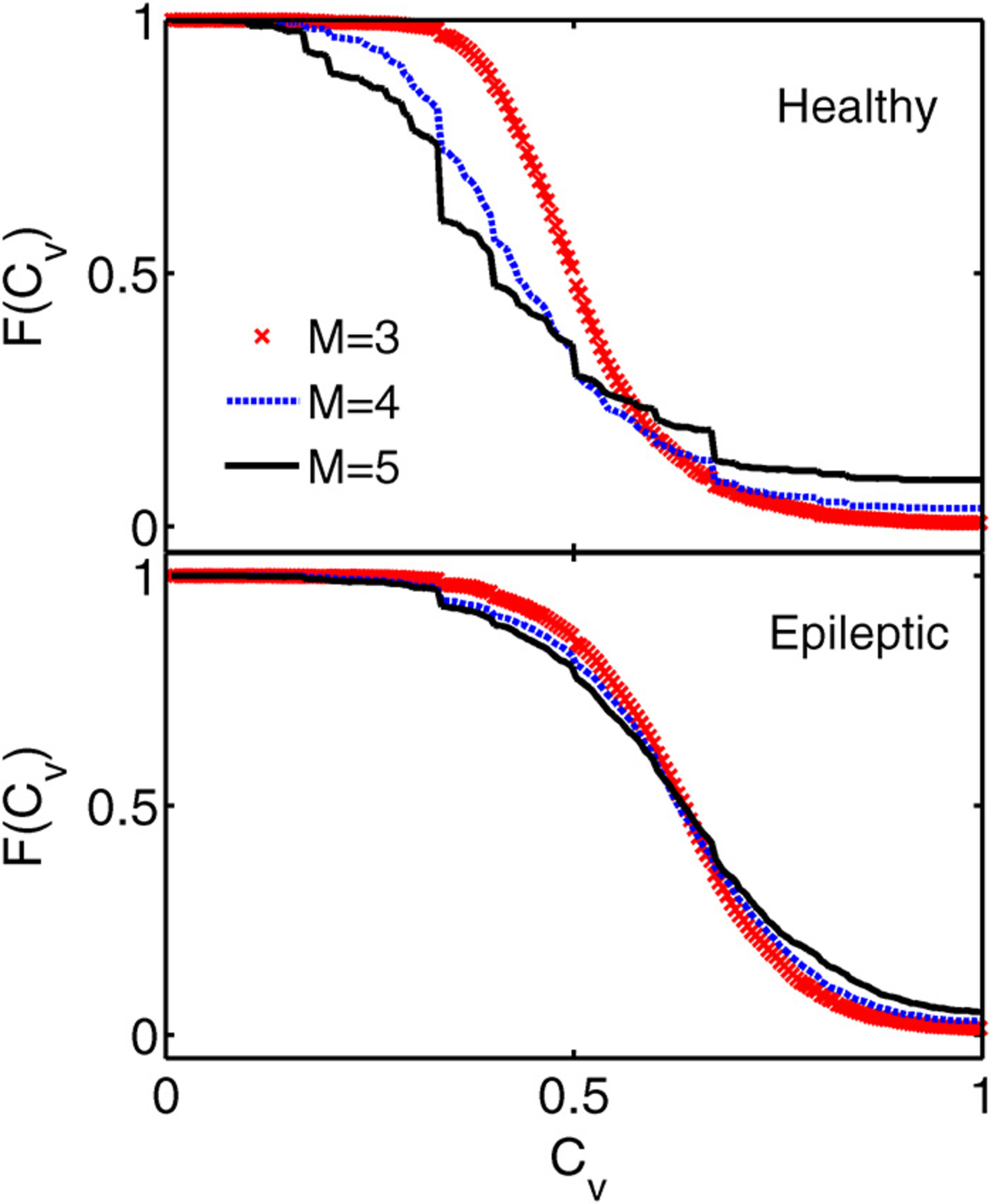}%
\caption{Cumulative distributions of the local CCs of RNs constructed from the EEG signals shown 
in the previous figure, with $M$ values 3, 4 and 5.} 
\label{f.7}
\end{figure}

Finally, as an example of application to real world data, we consider two typical EEG signals, one from a 
healthy person and the other from a person having epileptic seizure. Both data consist of $4098$ data 
points and are shown in Fig.~\ref{f.6}. The data have been studied earlier and further details regarding 
its generation and analysis can be found in Andrzejak et al. \cite {and}. There are indications of 
nonlinear signature in the dynamical properties of human brain's electrical activity, particularly 
during epileptic seizure, making the signal low dimensional. We now construct RNs from the two signals 
and compute the distributions $P(C_{\nu})$ for different embedding dimensions $M$. The cumulative 
distributions for $M = 3$, $4$ and $5$ for both signals are shown in Fig.~\ref{f.7}. 

\begin{figure}
\includegraphics[width=0.90\columnwidth]{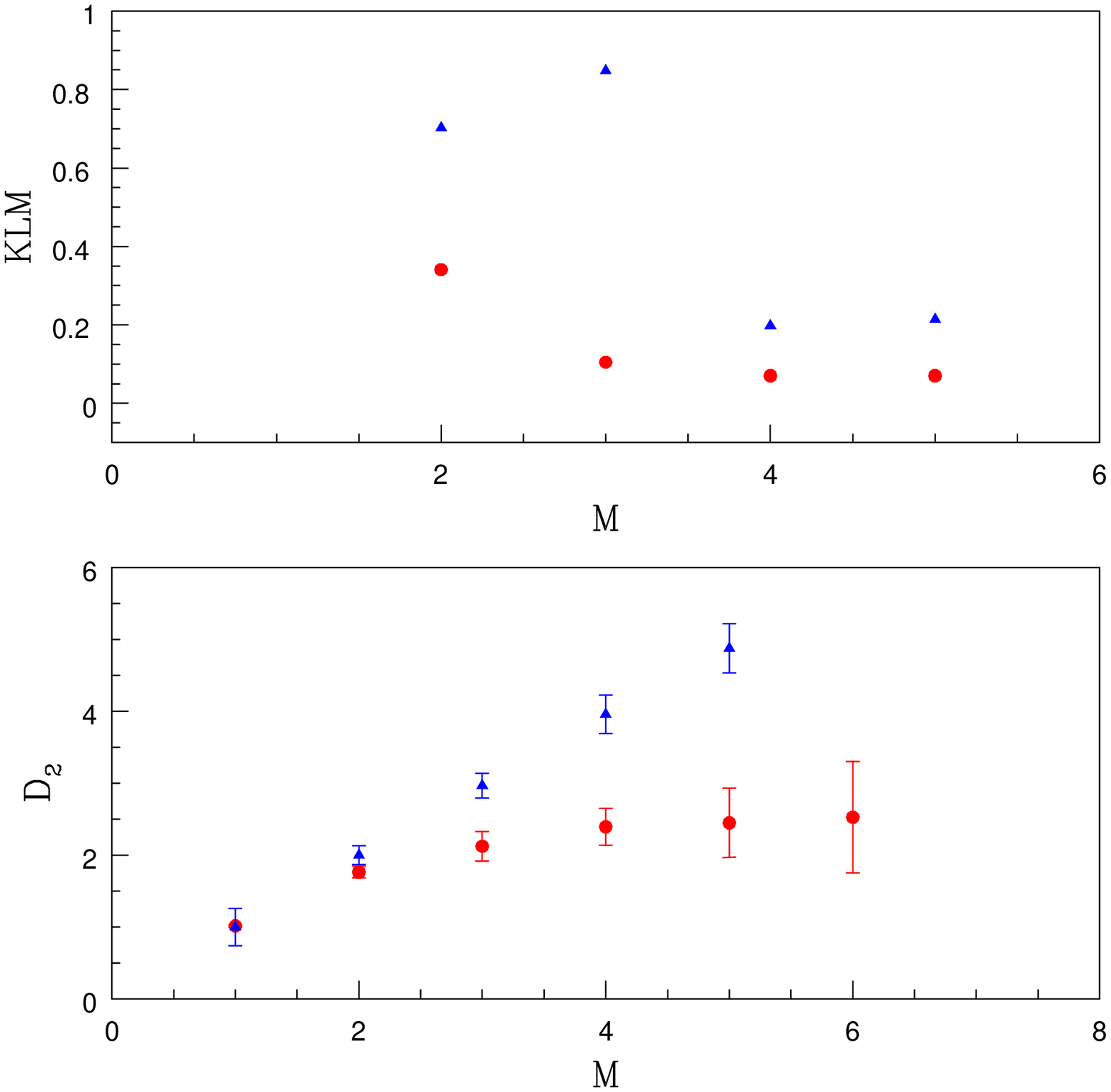}%
\caption{Variation of KLM with $M$ computed for the probability distributions of CCs from the 
two EEG signals (top panel). The solid triangle is for healthy and the solid circle is for 
the signal during seizure. The variation of $D_2$ with $M$ for the corresponding signals are 
shown in the bottom panel.} 
\label{f.8}
\end{figure}

The measure KLM for the distributions are computed as a function of $M$ for both signals and their 
variation is shown in Fig.~\ref{f.8} (top panel). It is clear that while the measure for the signal 
during seizure shows convergence for $M > 3$, such a tendency is absent for the healthy signal. 
This indicates that the latter is either high dimensional or involves fair amount of noise. To test 
this, we compute the value of correlation dimension $D_2$ as a function of $M$ for the two signals 
using the box counting scheme \cite {kph}. The results are also shown in Fig.~\ref{f.8} (bottom panel). 
As expected, the state of seizure shows nonlinearity with $D_2$ saturating at a low value $< 3$ 
while the $D_2$ for the healthy state keeps on increasing with $M$. This shows that the measure is 
fairly accurate in determining the embedding dimension for low dimensional signals.

\section{\label{sec:level1}Conclusion}
Network based measures are now increasingly being applied for the nonlinear analysis of time series data. 
An important advantage of such measures is the improved accuracy compared to conventional measures in the 
analysis of short and non-stationary data usually obtained from the real world. Here we present a RN 
based measure to determine the necessary embedding dimension to be used for the delay-embedding, a 
method normally used to reconstruct the dynamics in a higher dimension from the time series. 

The method involves computing the probability distributions of the local clustering coefficients of all 
nodes in the RNs constructed for successive embedding dimensions. A measure based on the K-L divergence 
is used to determine the convergence of the probability distributions and the value of $M$ where this 
happens is chosen as the required dimension for embedding. Eventhough the K-L divergence is a well 
known measure in information theory, its application in nonlinear time series analysis is novel. 

We show that the measure can accurately determine the dimension for standard low dimensional chaotic 
attractors and for data with small amount noise contamination $(< 20\%)$. To highlight the importance 
of the measure in the analysis of practical data, we apply it to two sample EEG signals, one from a 
healthy person and the other during epileptic seizure. 

The proposed method is analogous to the method of FNN. One may consider this as a RN approach to find 
the FNN. This is because, we can consider the nodes connected to a reference node in the RN as defining 
the number of nearest neighbors to a reference point on the attractor. Through local clustering 
coefficient, we are finding how many nodes connected to a reference node are also mutually connected. 
In other words, how many nearest neighbors to a point are also mutually nearest neighbors. Thus, for 
chaotic attractors, the number of nodes with a given local CC remains constant beyond the actual dimension. 
This, in turn, makes the probability distribution converge as a function of $M$. This convergence is 
quantified using a standard measure.

Finally, we do not claim that this measure by itself is accurate enough to give the proper embedding 
dimension for all the different types of data from the real world. For example, we have not checked 
the case of data with high dimension, say, $M > 4$. This may require time series with data length 
very large. Another possible issue is the amount of noise involved in the data about which, a priori, 
we have no knowledge. If the distribution does not show any convergence for the first few possible 
pair of $M$ values, we can only say that the data is either high dimensional or involves fair amount 
of noise. However, for low dimensional data with relatively low noise level, the method is 
simple and accurate even with limited data length.

\begin{acknowledgements}
RJ, KPH and RM acknowledge the financial support from the Science and Engineering Research  
Board (SERB), Govt. of India in the form of a Research Project No. SR/S2/HEP-27/2012.  
KPH acknowledges the computing facilities in IUCAA, Pune. 
\end{acknowledgements}

\end{document}